\documentclass[useAMS,usenatbib]{mn2e}

\usepackage{savesym}
\usepackage{graphicx}
\usepackage{fixltx2e} 
\usepackage{dcolumn}
\usepackage[version=3]{mhchem} 
\usepackage{natbib}
\usepackage{color}
\usepackage{latexsym}

\newcommand{\kms}{km\,s$^{-1}$}
\newcommand{\um}{${\rm \mu m}$}
\newcommand{\solarM}{M$_\odot$}
\newcommand{\solarL}{L$_\odot$}

\title[Infall, outflow, and turbulence in HMSF]{Infall, outflow, and turbulence in massive star-forming cores in the G333 Giant Molecular Cloud}

\author[N. Lo et al.]{N.~Lo$^{1}$, B.~Wiles$^{2}$, M. P.~Redman$^{2}$, M. R.~Cunningham$^{3}$, I.~Bains$^{4}$, P. A.~Jones$^{1,3}$, \and M. G.~Burton$^{3}$ and L.~Bronfman$^{1}$\\
  $^{1}$Departamento de Astronom\'ia, Universidad de Chile, Camino El Observatorio 1515, Las Condes, Santiago, Casilla 36-D, Chile\\
  $^{2}$Centre for Astronomy, School of Physics, National University of Ireland Galway, University Road, Galway, Ireland\\
  $^{3}$School of Physics, University of New South Wales, Sydney 2052, Australia\\
  $^{4}$Centre for Astrophysics and Supercomputing, Swinburne University of Technology, P.O. Box 218, Hawthorn, VIC 3122, Australia\\
}

\begin{document}

\date{DRAFT}

\pagerange{\pageref{firstpage}--\pageref{lastpage}} \pubyear{}

\maketitle

\label{firstpage}

\begin{abstract}
We present molecular line imaging observations of three massive molecular outflow sources, G333.6--0.2, G333.1--0.4, and G332.8--0.5, all of which also show evidence for infall, within the G333 giant molecular cloud (GMC). All three are within a beam size (36 arcseconds) of {\it IRAS} sources, 1.2-mm dust clumps, various masing species and radio continuum-detected H{\sc ii} regions and hence are associated with high-mass star formation. We present the molecular line data and derive the physical properties of the outflows including the mass, kinematics, and energetics and discuss the inferred characteristics of their driving sources. Outflow masses are of 10 to 40 \solarM in each lobe, with core masses of order $10^3$ \solarM. outflow size scales are a few tenth of a parsec, timescales are of several $\times 10^4$ years, mass loss rates a few $\times 10^{-4}$ \solarM/yr. We also find the cores are turbulent and highly supersonic.
\end{abstract}


\section{Introduction} \label{sec:intro}
The processes surrounding the life and death of massive stars play an important part in the evolution of galaxies at all epochs \citep[see e.g.][]{Hennebelle2012,Safranek_Shrader2014}, while at all times turbulence in the interstellar medium (ISM) plays a predominant role in regulating massive star formation \citep{Federrath2013}. Hence, understanding how turbulence shapes star formation, and how star formation in turn contributes to driving interstellar turbulence, is an important step in understanding diverse phenomena such as the evolution of the molecular ISM in galaxies, the formation of massive stars, and, eventually, the role that turbulence may play in the formation of planetary systems.

Three particularly important topics for understanding the interaction between turbulence and massive star formation are: i) The sources of the energy required to drive interstellar turbulence; ii) their relative importance at different scales \citep*[with large-scale Galactic flows, supernova explosions, outflows from young, massive stars and expanding H{\sc ii} regions all likely to contribute at various scales e.g.][]{Mac_Low2004, McKee2007}, and iii) The effect that turbulence and energy injection may have on enhancing or disrupting star formation at large, spiral-arm scales \citep{Luna2006} and smaller giant-molecular-cloud-size scales \citep[see e.g.][]{Harper_Clark2011}.

The study of infall and outflow in massive star forming regions is well connected to the subject area of turbulence. Outflow from massive stars may contribute to the driving of turbulence in the ISM \citep[e.g.][]{Rivilla2013, Federrath2014}, while turbulent fragmentation of gas that is infalling on to a protostellar cluster may change the number and mass distribution of the stars forming in the cluster \citep{Peters2010,Girichidis2012}.

Bipolar molecular outflows are found ubiquitously across all forming stellar size scales \citep[e.g.][]{Su2004,Wu2005,Zhang2013} down even to brown dwarfs \citep{whelan07}. The mass of molecular material observed in the outflows from high mass star forming regions (HMSFRs) is sufficiently large that it is likely entrained from the surrounding interstellar medium (ISM) in addition to the component associated with the forming star \citep*{Klaassen2008}. The driving mechanism for outflows from HMSFRs remains unclear but may also be due to a similar process to that of low mass star formation. In the accretion model of star formation \citep*[e.g.][]{Shu1987}, gravitational infall of the surrounding material onto a disc surrounding a forming star leads to mass ejection and the dissipation of excess angular momentum in the form of a bipolar jet; the causality of infall and outflow means where one is detected, the other is likely to be present. The \citet*{Shu1987} model considers isolated star formation; modelling suggests that outflows associated with high-mass stars can be highly collimated. In an interesting and recent development, \citet{MacLow2014} show that when multiple stars form in a common accretion flow, such as accretion onto a massive protostellar cluster, many star and protostars within the cluster have common outflow axes. Hence, even with massive star formation, there is likely to be a strong association between infall and outflow. Observational evidence for this scenario can be found in \citet*{Klaassen2008}. 

To provide benchmarking observational constraints for the role of turbulence in giant molecular clouds, a multi-molecular line mapping of the G333 Giant Molecular Cloud (GMC) with the Mopra and Nanten2 telescopes has been undertaken \citep[see][]{Bains2006, Wong2008, Lo2007, Lo2009, Lo2011}. In collecting data for this project, we have serendipitously detected signatures of outflow and infall in three of the brighter molecular features in the GMC, which we have designated G333.6--0.2, G333.1--0.4 and G332.8--0.5. Within a Mopra beam size of the outflow sources are radio-detected H{\sc ii} regions, 1.2-mm dust emission clumps \citep{Mookerjea2004}, \ce{CH3OH} and \ce{H2O} masers and {\it IRAS} sources, features which are all consistent with the presence of high mass star formation. 

\subsection{G333 outflow/infall sources} \label{sec:outin}
\begin{figure*}
  \includegraphics[]{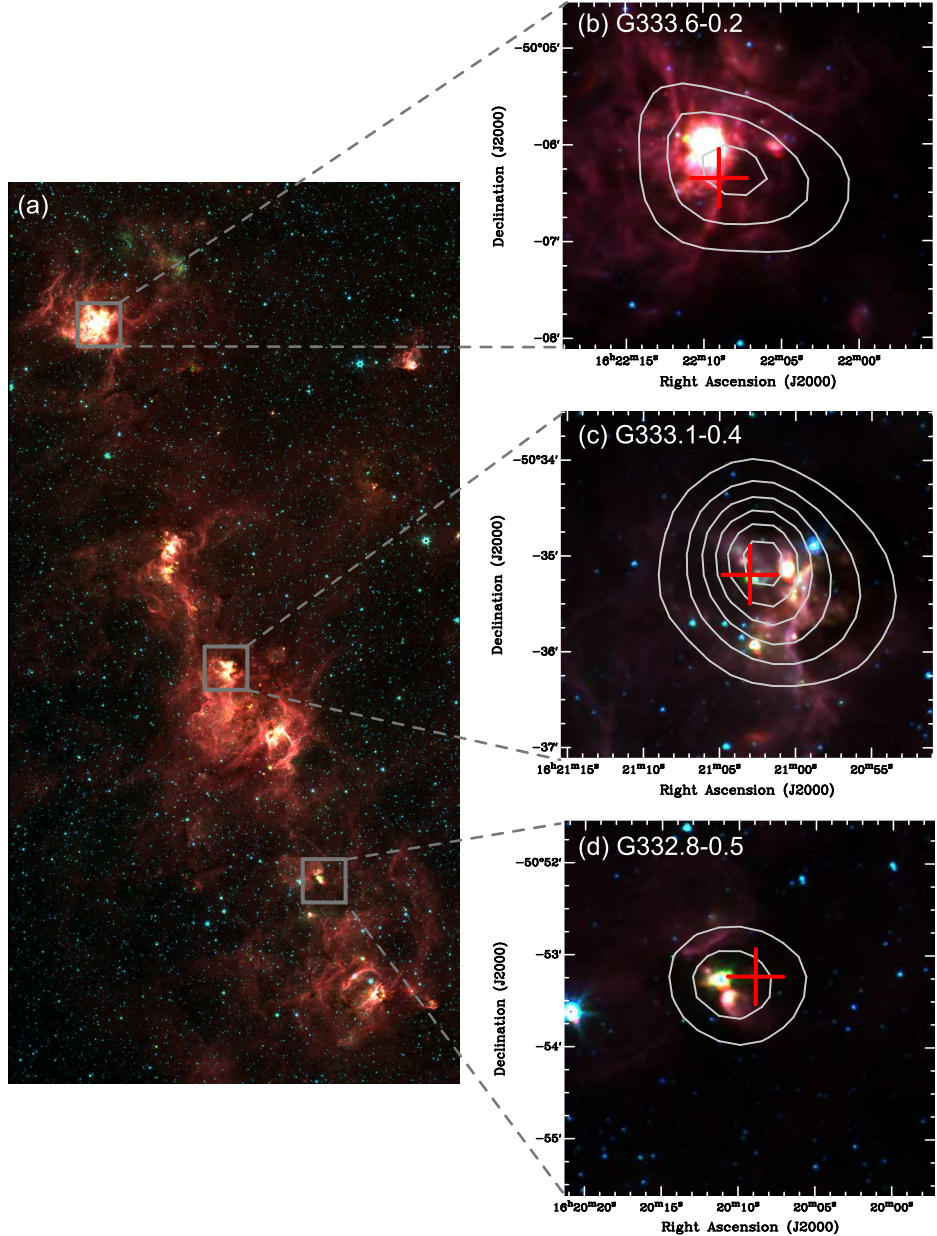}
  \caption{Colour-composite images of the {\it Spitzer} IRAC photometric bands for the vicinity of the outflow sources where red is 8.0-\um, green is 4.5-\um\ and blue is 3.6-\um. Overlain are white contours of total intensity Mopra \ce{CS} ($J=2-1$) emission, which picks out the cores of the outflow sources. Contour levels start at 10 K \kms, with increments of 5 K \kms\ up to 35 K \kms. {\it (a)} Overall image of the G333 region with the positions of the three sources in this paper indicated. Zoom in image of the three regions: {\it (b)} G333.6--0.2, {\it (c)} G333.1--0.4 and {\it (d)} G332.8--0.5. The crosses mark the positions for the spectra shown in Figure~\ref{fig:profiles}, as listed in Table~\ref{tab:outflowproperties}, which are also the peaks of \ce{CO} emission. The angular separations between the 8-\um\ and \ce{CO} peaks are 20, 15 and 7 arcseconds for G333.6--0.2, G333.1--0.4 and G332.8--0.5 respectively, note that the beam size of the Mopra data is 36 arcseconds, thus all within one bean size. (Colour image in the electronic version.)}
  \label{fig:glimpse}
\end{figure*}

In this paper we examine evidence of infall and outflow towards three massive star forming sources in the G333 massive molecular cloud complex (see Figure \ref{fig:glimpse}). The {\bf G333.6--0.2} molecular outflow is associated with the most massive of the three HMSFRs and harbours a young OB cluster \citep[e.g.][]{Fujiyoshi2006}, with a dust mass of $1.6 \times 10^4$ \solarM \citep{Mookerjea2004}. The {\it IRAS} source closest to G333.6--0.2, IRAS 16183--4958, is one of the most luminous far-infrared (FIR) sources in the sky. From observations with the VLT MIR VISIR camera, \citet{Grave2014} found indications that this region consists of two main luminous sources (O4V and O5V) which account for at least half of the luminosity from this region. As well as being proximate to the H{\sc ii} region G333.6--0.22, this molecular outflow source is also associated with {\it MSX} and 1.2-mm dust sources \citep{Mookerjea2004} and masers of \ce{H2O} and \ce{OH} \citep{Batchelor1980,Caswell1998}.

The HMSFR associated with {\bf G333.1--0.4} is intermediate in mass to the other two sources discussed in this paper. High resolution deep near-IR imaging and spectroscopy reveals an embedded OB star cluster in very early evolutionary stages \citep{Figueredo2005}. Spectroscopy shows the two brightest stars in the cluster have spectral type of O6 and O8 stars, and numerous young stellar objects with excess near-IR emission due to circumstellar disks or envelopes. The cluster has integrated mass of $1.0 \times 10^3$ \solarM. This source also has the most prominent \ce{SiO} emission in the whole G333 complex, aside from the cold core G333.125--0.562 as discussed in \citet{Lo2007}.

Finally, we also observed {\bf G332.8--0.5} which is the smallest of the three HMSFRs, with a dust mass of $5.5 \times 10^3$ \solarM \citep{Mookerjea2004}. It has FIR colour characteristics of an ultra compact H{\sc ii} region \citep{Bronfman1996}. Table~\ref{tab:assocsources} gives a summary of previous observational identifications of the three sources in infrared continuum and maser transitions.

The molecules presented in this paper trace different conditions in the interstellar medium related to infall, outflow and the dense cores in the sources. Being such a ubiquitous species, the \ce{^{12}CO} emission traces the full spatial and kinematical extent of the outflows. The \ce{^{13}CO} traces them to a lesser extent, but where present it can be used with the \ce{^{12}CO} to calculate the molecular column density (under the assumption that the \ce{^{13}CO} is optically thin). Emission from \ce{C^{18}O} is mainly confined to the cloud cores and provides a measure of the core mass, if assumed to be optically thin. \ce{SiO} emission is known to be enhanced in outflows due to the presence of shocks, although it is not found in every outflow source \citep*{Klaassen2007}. \ce{HCO+} can trace both infall and outflows \citep[e.g.][]{Rawlings2004,Myers1996,Klaassen2007} as it easily becomes optically thick and traces a high critical density. Emission from the \ce{SiO} and \ce{HCO+} lines therefore provides a tracer of outflow and infall phenomena. \ce{CS} is a high density ($\rm \sim 10^5 cm^{-3}$) tracer which is found towards star-forming condensations rather than outflow wings and so traces the systemic velocity of the clouds and the degree of turbulence. \ce{N_2H+} is prominent in cold, dense cloud cores rather than in outflow wings, typically at $T_{\rm ex} \la 20 \, K$, due to its main destroyer \ce{CO} being depleted \citep*{Bergin1995}. 

For the three G333 sources, we present Mopra data of molecular line emission which show evidence for outflow and infall, specifically that of the low excitation rotational transitions of three \ce{CO} isotopologues, \ce{CS}, two isoptopologues of \ce{HCO+}, \ce{SiO} and \ce{N_2H+}. The observations are described in Section \ref{sec:observations}. In Section \ref{sec:results}, we present archival {\it Spitzer} GLIMPSE mid-IR imagery of the three outflow sources overlaid with contours of Mopra \ce{CS} ($J=2-1$) data, followed by the Mopra spectral line profiles and discuss the evidence for the presence of infall from these lines. In Section \ref{sec:analysis}, we use the CO isotopologue data to calculate the column density in the outflows and use this to derive their mass and energetics. In Section \ref{sec:summary} we summarise the observationally derived properties we have determined for these sources and introduce a companion paper \citep{Wiles2015} in which models of the regions are presented.

\section{OBSERVATIONS AND DATA REDUCTION} \label{sec:observations}

\begin{table*} 
  \centering
  \caption{Names associated with continuum and maser sources found within 80 arcseconds from the sources from the {\sc SIMBAD} database. The Mopra beam is 36 arcseconds at 3-mm wavelengths. The references are: (a) \citet{Mookerjea2004}; (b) \citet{Becklin1973}; (c) \citet{Murphy2010}; (d) \citet{Figueredo2005}; (e) \citet{Breen2007}; (f) \citet*{Caswell1998}; (g) \citet{Caswell1995}.}
  \label{tab:assocsources} 
  \tiny
  \begin{tabular}{@{}ccccccccc} 
    \hline
    Source & RA, Dec & IRAS & MSX & 1.2-mm$\rm ^a$ & H{\sc ii} &\ce{H2O}$\rm ^e$ & \ce{OH}$\rm ^f$ & \ce{CH3OH}$\rm ^g$ \\
     & (J2000) &&&&&&& \\
    \hline
    G333.6--0.2 & 16:22:09.0, -50:06:21 & IRAS 16183-4958 & G333.6046-00.2124 & MMS5 & G333.60-00.21$\rm ^b$ & G333.608−0.215 & G333.608-00.215 & - \\
     & & \ldots & G333.6044-00.2165 \\
    G333.1--0.4 & 16:21:03.3, -50:35:12 & IRAS 16172-5028 & G333.1104-00.4223 & MMS39 & G333.1-00.4$\rm ^d$ & G333.121−0.434 &  G333.135-00.431 & G333.121-00.434 \\
     & & \ldots & G333.1306-00.4257 & MMS40 \\
    G332.8--0.5 & 16:20:08.9, -50:53:14 & IRAS 16164-5046 & G332.8269-00.5489 & MMS68 & G332.8-00.6$\rm ^c$ & G332.826−0.549 & G332.824-00.548 & - \\
    \hline 
  \end{tabular}
\end{table*}

\begin{table*}
  \centering
  \caption{Molecular lines observed. Columns are as follows: (1) molecule, (2) transition; (3) rest frequency (for \ce{N2H+} only the frequency of the main hyperfine line is given); (4) months for observations; (5) velocity resolution; (6) typical $1 \sigma$ rms off-line noise level per velocity channel in terms of measured $T_{\rm A}^*$; (7) reference.}
  \label{tab:obsdetails}
  \begin{tabular}{@{}lcccccc}
    \hline
    Molecule & Transition & Rest frequency & Months Observed & $\Delta v$ & $1 \sigma$ & Reference \\
     &	       & GHz	    &	   &	\kms	  &	K	   &	\\
    (1)	 &	(2)    & (3)	    &	(4)	   &	(5)	  &	 (6)   &	(7) \\
    \hline
    \ce{^{12}CO} &	$J=1-0$ & 115.27 &	2006 Aug	   &	0.09	  &	0.2	 & this work \\
    \ce{^{13}CO} &	$J=1-0$ & 110.20  &	2004 Jun - Oct &	0.17	  &	0.1	 & \citet{Bains2006} \\
    \ce{C^{18}O} &	$J=1-0$ & 109.78 &	2005 Jul - Sep &	0.17	  &	0.1	 & \citet{Wong2008} \\
    \ce{CS}		 &	$J=2-1$ &  97.981 &	2006 Sep - Oct &	0.10	  &	0.1	 & \citet{Lo2009} \\
    \ce{HCO+}	 &	$J=1-0$ &  89.190 &	2006 Jul - Sep &	0.11	  &	0.1	 & \citet{Lo2009} \\
    \ce{H^{13}CO+} &$J=1-0$ &  86.754 &	2006 Jul - Sep &	0.12	  &	0.2	 & \citet{Lo2009} \\
    \ce{N2H+} &	$J=1-0$ &  93.173    &	2006 Sep - Oct &	0.11	  &	0.1	 & \citet{Lo2009} \\
    \ce{SiO} & $J=2-1$ &  86.847 &	2006 Jul - Sep &	0.12	  &	0.1	 & \citet{Lo2009} \\
    \hline
  \end{tabular}
\end{table*}

The molecular line data presented here are comprised of data from our Mopra G333 multi-molecular lines mapping \citep{Bains2006,Wong2008,Lo2009} as well as \ce{^{12}CO} maps of the individual sources. For a more detailed description of the observing procedure and data reduction steps, we refer the reader to the references mentioned immediately above.

The Mopra\footnote{The Mopra radio telescope is part of the Australia Telescope National Facility which is funded by the Commonwealth of Australia for operation as a National Facility managed by CSIRO.} radio telescope is a 22-metre-single-dish telescope located near Coonabarrabran, NSW, Australia, with a beam size of 36 arcseconds at 3-mm wavelengths. The main beam brightness temperature $T_{\rm MB}$ and antenna temperature $T_{\rm A}^*$ are related by the antenna efficiency $\eta_{\nu}$ at frequency $\nu$ such that $T_{\rm MB} = T_{\rm A}^* / \eta_{\nu}$ and this was used to derive $T_{\rm MB}$. The Mopra beam has been characterised by \citet{Ladd2005} and the beam efficiencies used are as listed there. The observing bandwidth was configured so that the central channel corresponded to $-50$ \kms, the approximate velocity at which the emission from the GMC complex is centred. The reference (OFF) position is at $\alpha_{\rm J2000} = 16:27$, $\delta_{\rm J2000} = -51:30$ \citep{Bains2006}. Throughout this work, velocities are given in the radio convention and in terms of $v_{\rm lsr}$, that is, with respect to the kinematic local standard of rest (lsr). The data were reduced using the {\sc livedata} and {\sc gridzilla} packages available from the CSIRO/CASS\footnote{See URL http://www.atnf.csiro.au/computing/software/livedata }, weighted by the relevant $T_{\rm sys}$ measurements, and have been continuum subtracted. We summarise the observational details in Table~\ref{tab:obsdetails}. The G333 cloud was observed in a number of other species in addition to those listed in section 1.1, as detailed in \citet{Lo2009}; however, these are not discussed in this paper.

\section{RESULTS} \label{sec:results}
\subsection{Mopra Molecular Line Data} \label{sec:mopramoldata}
\subsubsection{Velocity Profiles} \label{sec:velprof}

\begin{figure*}
  \includegraphics[width=1.0\textwidth]{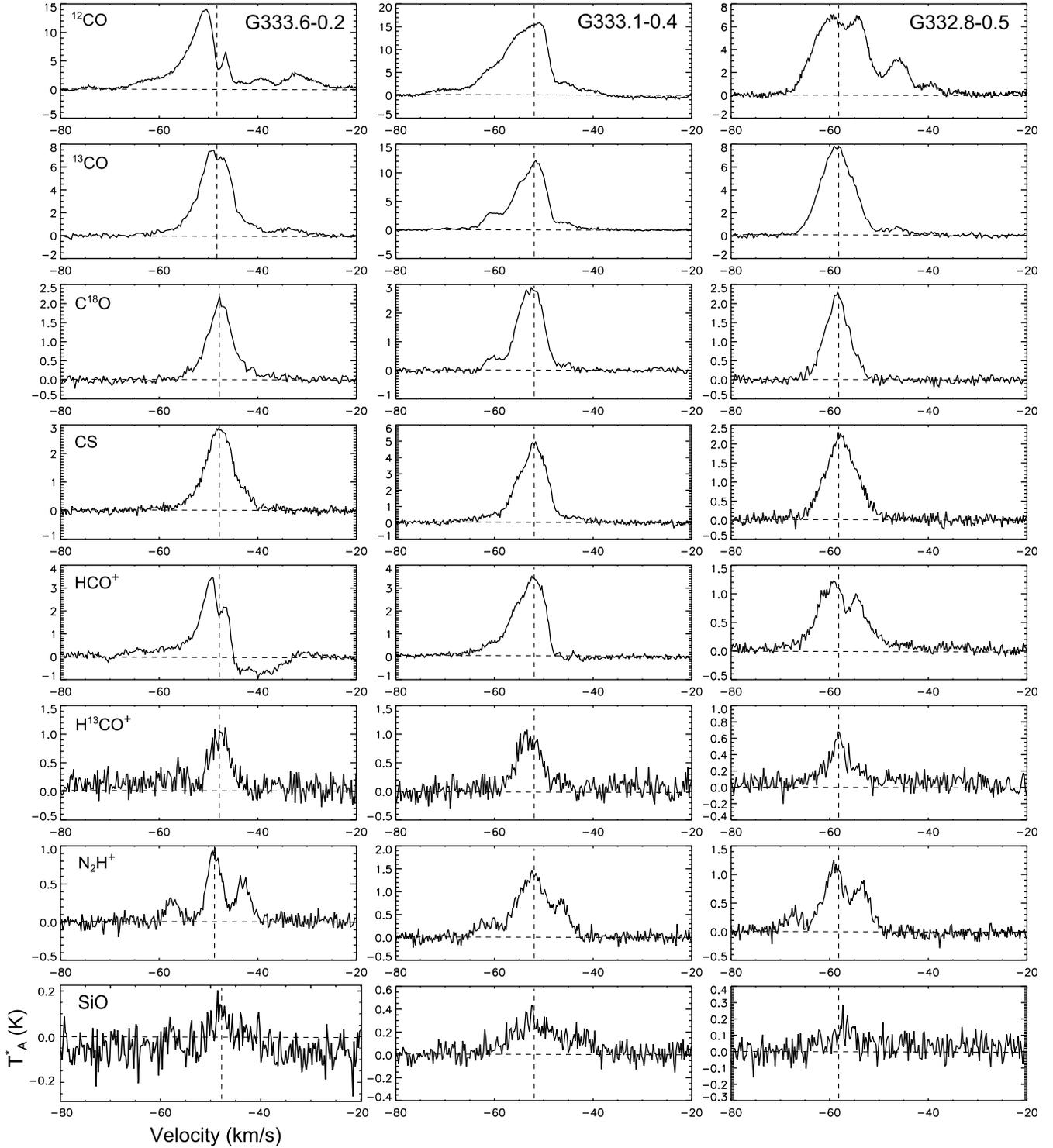}
  \caption{Velocity profiles of the molecular lines for each of the outflow sources. The spectra were taken from the positions indicated by the red crosses in Figure \ref{fig:glimpse} (as listed in Table~\ref{tab:outflowproperties}). Left: G333.6--0.2; centre: G333.1--0.4; right: G332.8--0.5.  From top to bottom the lines are (see Table~\ref{tab:obsdetails}) \ce{^{12}CO}, \ce{^{13}CO}, \ce{C^{18}O}, CS, \ce{HCO+}, \ce{H^{13}CO+}, \ce{N_2H+} and SiO. The y-axis is the intensity, $T_{\rm A}^*$, in units of K and the x-axis is the $v_{\rm lsr}$ velocity in \kms. The dashed lines indicate $T_{\rm A}^* = 0$ (horizontal) and centroid velocity (vertical).}
  \label{fig:profiles}
\end{figure*}

In Figure~\ref{fig:profiles}, we show the molecular line velocity profiles taken at the spatial location of the peak \ce{CO} emission for each source (marked with red crosses in Figure \ref{fig:glimpse} and listed in Table~\ref{tab:outflowproperties}). The displayed velocity range of the profiles was determined from the \ce{^{12}CO} data, the line which traces the maximum extent of the outflows; it is this velocity range that was used to perform the analysis of the outflows as described in Section \ref{sec:analysis}.

In each source, multiple velocity features are visible in some line profiles, particularly those of \ce{^{12}CO} and \ce{HCO+}, whilst \ce{CS}, \ce{C^{18}O} and \ce{H^{13}CO+} show a single velocity feature. \ce{CS} is a high-density tracer and thus associated with the denser cloud core region, while the \ce{H^{13}CO+} isotopologue is assumed to be optically thin and hence also traces the core rather than the outflow. The \ce{C^{18}O} emission also appears to mainly trace the cloud core component. \ce{N_2H+} is found in cold dense cores rather than outflows. In the case of G333.6--0.2 and G332.8--0.5, we consider these to be associated with a single source rather than several overlapping velocity features along the line-of-sight. This is evidenced by the \ce{CS}, \ce{C^{18}O} and \ce{H^{13}CO+} line profiles, which comprise a single component at the same central velocity. 

In both G333.6--0.2 and G332.8--0.5, the \ce{^{12}CO} and \ce{HCO+} lines all show the broad wings characteristic of outflows, particularly on the blue-shifted side in the case of G333.6--0.2 (detailed analysis see Section \ref{sec:mol_in_out_turb}). In G333.1--0.4, a blue-shifted shoulder (at $\sim -60$ \kms) is present in all three \ce{CO} isotopologues and also \ce{HCO+}; the fact that this feature is present in both optically thick (\ce{CO}) and thin (\ce{C^{18}O}) lines, unlike the other two sources, may suggest that it is due to a confusing cloud along the same line-of-sight rather than part of the wing emission in G333.1--0.4. The detection of emission from \ce{HCO+} and \ce{SiO} in all the sources is consistent with the presence of outflows \citep{Rawlings2004,Klaassen2007}.

\begin{figure}
\includegraphics{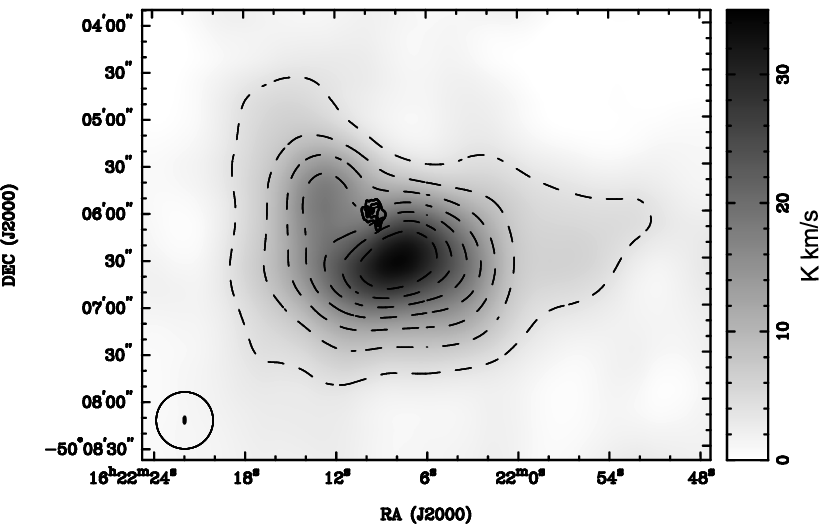}
  \caption{Total intensity of \ce{HCO+} greyscale (Mopra Telescope) with dashed contours, and black contours overlaid from the G333.6--0.22 H{\sc ii} region at 4.8-GHz (ATCA). The lower left circles indicate the beam sizes of the data, 36 and 6 arcseconds for Mopra and ATCA respectively.}
  \label{fig:g333_6_hcop_cont5GHz}
\end{figure}

An infall signature in a spectral line presents itself in the form of a red-blue asymmetry, usually with a diminished red-shifted component \citep[e.g.][]{Walker1994, Myers1996}. Such an asymmetry is apparent to varying degrees in the line profiles of all three sources. G333.6--0.2 shows the most extreme and broadest red-shifted infall feature, extending from $\sim -33$ \kms\ bluewards towards the centre of the line in \ce{^{12}CO} and \ce{HCO+} (Figure~\ref{fig:profiles}) and also visible to a lesser degree in \ce{^{13}CO}. Continuum absorption has further `distorted' the line shape of \ce{HCO+} such that the infall feature is negative (the profiles have been continuum subtracted). This is consistent with the presence of a number of radio- and millimetre-wavelength detected continuum sources within G333.6--0.2 \citep[][see also Figure~\ref{fig:glimpse}]{Fujiyoshi2006}. Indeed, the peak in the continuum absorption of the molecular line emission occurs along the line-of-sight to the position of peak radio flux density to the H{\sc ii} region G333.6--0.22 (observed with the Australia Telescope Compact Array), which also coincides with the 1.2-mm dust peak (Figure~\ref{fig:g333_6_hcop_cont5GHz}).

G333.1--0.4 also clearly shows an abrupt fall-off on the red-shifted side of its line profiles (Figure \ref{fig:profiles}). This is a less extreme infall feature than that seen in G333.6--0.2 and is again consistent with the line being absorbed by the continuum source which we have detected at radio and millimetre wavelengths (see also Figure~\ref{fig:glimpse}). G332.8--0.5 shows a more `classic' infall profile, with a clear split of $\sim 5$ \kms\ between asymmetric red- and blue-shifted peaks ($-54$ and $-60$ \kms\ respectively) which is particularly evident in the line profiles of \ce{^{12}CO} and \ce{HCO+}.

\subsubsection{Outflow Maps} \label{sec:cochannel}
\begin{figure*}
  \includegraphics{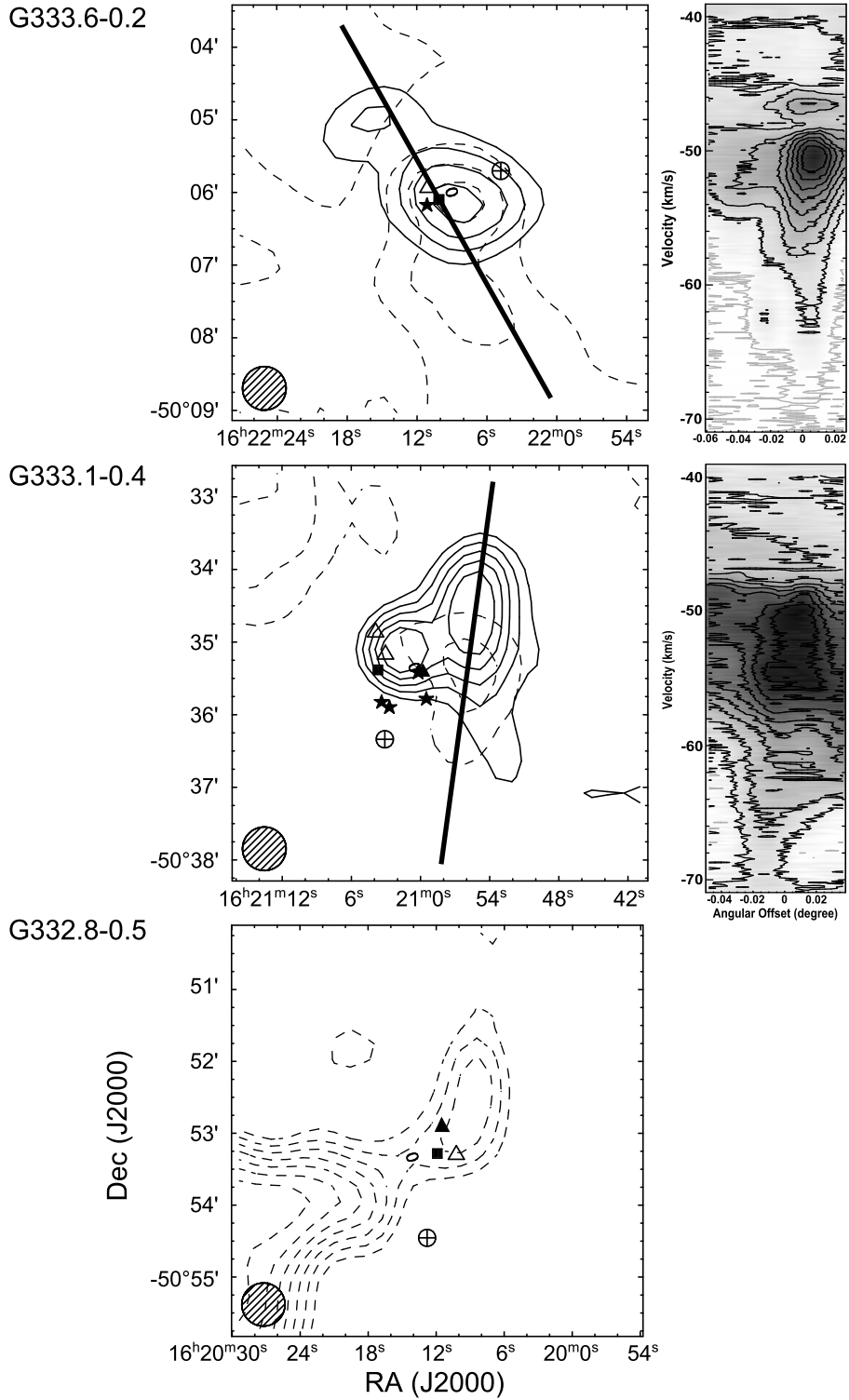}
  \caption{(left:) Total intensity \ce{^{12}CO} red- (dashed contours) and blue-shifted (solid) wing emission, summed over the velocity ranges given in Table~\ref{tab:outflowproperties}. For G333.6--0.2, contours are plotted in 5 K \kms\ steps starting at 5 K \kms\ for both blue and red shifted outflow emissions. For G333.1--0.4, contours are plotted in 2 K \kms\ steps starting at 14 K \kms\ for the blue-shifted emission, and in 1 K \kms\ steps starting at 8 K \kms\ for the red-shifted emission. For G332.8--0.5, there is no detectable blue-shifted emission, the red-shifted emission contours are in steps of 2 K \kms\ starting at 18 K \kms. The thick solid line on each plot indicates the position of the slice taken to produce the PV-arrays shown. (right:) Position-velocity arrays with contours plotted at 10 per cent steps of the peak emission. (top:) G333.6--0.2. PV-array centred on RA 16:22:9.2, Dec -50:06:02, PA 34$^{\circ}$. (middle:) G333.1--0.4. PV-array centred on RA 16:21:0.6, Dec -50:35:16, PA 351$^{\circ}$. (bottom:) G332.8--0.5. PV-array is not shown as the PA is unknown. All angles are measured E from N. The symbols indicate the other sources of emission in the region (see Table~\ref{tab:assocsources}) as follows: {\it MSX}--box, \ce{H2O} maser--filled triangle, OH maser--$\bigtriangleup$, \ce{CH3OH} maser--$\star$, H{\sc ii} region--$\oplus$, 1.2-mm source--$\ast$ and the {\it IRAS} sources are indicated by their error ellipses.}
  \label{fig:outflows_br} 
\end{figure*}

In Figure~\ref{fig:outflows_br}, we present images of the \ce{^{13}CO} data to show the overall velocity structure.  The zeroth moment (i.e. total intensity) contours of the red and blue wing emission (summed over the velocity ranges defined in Table~\ref{tab:outflowproperties}) for each outflow source are shown overlaid with symbols indicating the positions of other likely related sources of emission in the region, more details of which are given in Table~\ref{tab:assocsources}. Two of the three sources (G333.2--0.6 and G333.1--0.4) have a clear offset between the red- and blue-shifted total intensity emission; the position angle (PA) of this offset is different for each source, i.e. there is no evidence for a general alignment of such kinematic characteristics across the large-scale G333 GMC. However, care must be taken in the interpretation of the red-shifted total intensity emission from G333.1--0.4 and G333.6--0.2 because of the red-shifted absorption dip in each of these sources. For source G332.8--0.5, there is no measurable blue-shifted emission thus it is not possible to determine the outflow axis.

We also show the PV (position velocity)-arrays in the right column of Figure~\ref{fig:outflows_br} , made by taking a slice through the \ce{^{12}CO} data cube along a plane defined by the positions of the peak emission in the most extreme red- and blue-shifted channels; the slice positions are indicated by the thick solid lines overlaid on the moment images. The PV-arrays for G333.6--0.2 and G333.1--0.4 show the outflows blending into spatially and/or kinematically adjacent emission, and the abrupt falls in the red-shifted channel emission. The G332.8--0.5 PV-array is not shown because the outflow axis is not determined. 

\subsection{{\it Spitzer} GLIMPSE Imagery} \label{sec:glimpse}
In Figure~\ref{fig:glimpse}, we show the {\it Spitzer Space Telescope} GLIMPSE \citep[Galactic Legacy Infrared Mid-Plane Survey Extraordinaire;][]{Benjamin2003} 3-colour composite infrared images of the whole cloud complex and the three bipolar outflow sources. The images are overlain with contours of the \ce{CS} total intensity, which we display here as it has a high associated critical density ($\rm \sim 10^5 cm^{-3}$) and so picks out the structure of the high-density molecular cores well.

Infrared emission associated with all three outflow sources is apparent in the GLIMPSE images. G333.6--0.2 and G333.1--0.4 display complex structures comprising a number of knots and filaments in the IR while G332.8--0.5 has a simpler, bipolar nebulosity. The predominance of 8-\um\ emission (red) in all three sources is consistent with the presence of polycyclic aromatic hydrocarbons (PAHs) typically found in photodissociation regions (PDRs) around HMSFRs, so the 8-\um\ peak is taken as being a sign of an H{\sc ii} region. The positions of the H{\sc ii} regions visible in the {\it Spitzer} data are consistent with the positions of the radio continuum emission \citep{Urquhart2007}. Emission associated with the massive OB cluster in G333.6--0.2 \citep[e.g.][]{Fujiyoshi2006} is clearly visible as the saturated region in the {\it Spitzer} InfraRed Array Camera (IRAC) image and is clearly offset from the molecular emission peak, which is itself associated with a dark region in the IR emission. Similarly, the molecular peak in G333.1--0.4 is offset from the emission associated with the H{\sc ii} region and coincident with a trough in the IR emission. Such offsets between molecular emission and H{\sc ii} regions are consistent with more evolved HMSFRs whose feedback effects have cleared their natal environs. Conversely, the H{\sc ii} region in G332.8--0.5 is located in the centre of the peak molecular contour, suggesting it may be a younger source and/or less powerful.

\section{ANALYSIS} \label{sec:analysis}
\subsection{Molecular Lines: optical depths} \label{sec:mol_opacity}

As the three HMSFRs are located within a massive GMC complex, the analysis was hampered by the multiplicity of sources present and their associated confusion, both spatially and kinematically. The outflows are thus less well-defined than in low mass star forming regions (LMSFR) and the clean, detailed analysis that is possible there is not possible here.  

\begin{table}
  \caption{Gaussian fits to the \ce{C^{18}O}, \ce{^{13}CO} and \ce{CS} line profiles of the cloud core emission in each source (1). Both the centroid (2), (3), (4) and FWHM (5), (6), (7) velocities are given for each line.}
  \label{tab:gaussfits}   
  \begin{tabular}{@{}ccccccc} \hline
    Source	& \multicolumn{3}{c}{$v_{\rm lsr}$ (\kms)} & \multicolumn{3}{c}{FWHM (\kms)} \\
     & \ce{C^{18}O} & \ce{^{13}CO}	& \ce{CS}	& \ce{C^{18}O}	& \ce{^{13}CO}	& \ce{CS} \\	
    (1) & (2) & (3) & (4) & (5) & (6) & (7) \\
    \hline					
    G333.6--0.2 &	-47.7	& -48.6	& -47.7	& 6.1	& 7.5	& 7.0 \\
    G333.1--0.4 &	-52.7	& -52.6	& -52.2	& 6.1	& 7.4	& 6.8 \\
    G332.8--0.5 &	-58.5	& -58.3	& -57.7	& 5.9	& 7.7	& 7.0 \\
    \hline 
  \end{tabular} 
\end{table} 

\begin{table*}
  \caption{Line-derived excitation temperatures, isotopologue intensity ratios and optical depths. The excitation temperature  $T_{\rm ex}$ (2) is determined from the brightest value of \ce{^{12}CO} $T_{\rm A}^*$ found in each source (1), divided by $\eta_{\nu}$, at the $v_{\rm lsr}$ velocity listed (3). The maximum value for the isotopologue ratios \ce{^{12}CO}/\ce{^{13}CO} (4) and \ce{^{12}CO}/\ce{C^{18}O} (6) are also listed, together with the corresponding $v_{\rm lsr}$ velocities (5), (7) where this occurs. The maximum value for the optical depth (8) is determined from the minimum value of the \ce{^{12}CO}/\ce{C^{18}O} ratio, and occurs at the velocity (9) indicated.}
  \label{tab:lineparams}   
  \begin{tabular}{@{}ccccccccc}
    \hline
    Source & $T_{\rm ex}$ & $v_{\rm lsr}$ & \ce{^{12}CO}/\ce{^{13}CO}$_{\rm max}$ & $v_{\rm lsr}$ & \ce{^{12}CO}/\ce{C^{18}O}$_{\rm max}$ & $v_{\rm lsr}$ & $\tau_{\rm max}$ & $v_{\rm lsr}$ \\
     & K  & \kms &  & \kms  &  & \kms &  & \kms \\
    (1) & (2) & (3) & (4) & (5) & (6) & (7) & (8) & (9) \\
    \hline
    G333.6--0.2 &	30 &	-50.4 &	6.4 &	-55.7 &	25.2 &	-52.2 &	73 &	-44.7 \\
    G333.1--0.4 &	33 &	-50.9 &	4.4 &	-62.9 &	24.8 &	-57.4 &	20 &	-53.6 \\
    G332.8--0.5 &	17 &	-59.7 &	4.3 &	-51.6 &	13.8 &	-53.3 &	37 &	-58.3 \\
    \hline 
  \end{tabular} 
\end{table*} 

\begin{table*}
  \caption{Measured outflow parameters. The columns are as follows: (1) source; (2) \& (3) RA and Dec (J2000) giving the position of spectra shown in Figure~\ref{fig:profiles}; (4) \& (5) $v_{\rm B_1}$ \& $v_{\rm B_2}$ giving velocity limits for blue-shifted outflow; (6) \& (7) $v_{\rm R_1}$ \& $v_{\rm R_2}$ velocity limits for red-shifted outflows; (8) \& (9) $v_{\rm B_{mean}}$ \& $v_{\rm R_{mean}}$ the mean projected velocity of blue- and red-shifted outflows; (10) position angle for the outflow in the plane of the sky, measured E from N. Outflow velocities are determined after subtracting the scaled Gaussian fits to the \ce{C^{18}O} line from the extinction-corrected \ce{^{12}CO} line profile (see text) (note: no blue-shifted emission is detectable in the line profile for G332.8--0.5). The line core emission is taken as between $v_{\rm B_2}$ and $v_{\rm R_1}$. Parameters calculated for the core and outflow given in other Tables use these velocity limits, together with the line centre velocities given by the fit to the \ce{C^{18}O} profile (see Table~\ref{tab:gaussfits}).}
  \label{tab:outflowproperties}
  \begin{tabular}{@{}cccccccccc}
  \hline
    Source	& RA & Dec & $v_{\rm B_1}$ & $v_{\rm B_2}$ & $v_{\rm R_1}$ & $v_{\rm R_2}$ & $v_{\rm B_{mean}}$ & $v_{\rm R_{mean}}$ & Position \\
     & \multicolumn{2}{c}{J2000} & \kms & \kms & \kms & \kms & \kms & \kms & Angle ($^{\circ}$E of N) \\
    (1) & (2) & (3) & (4) & (5) & (6) & (7) & (8) & (9) & (10) \\
    \hline
    G333.6--0.2 &	16:22:09.0 &	-50:06:21 &	-69 &	-57 &	-37 &	-26 &	15.3 &	16.2 &	34 \\
    G333.1--0.4 &	16:21:03.3 &	-50:35:12 &	-75 &	-63 &	-43 &	-37 &	16.3 &	12.7 &	351 \\
    G332.8--0.5 &	16:20:08.9 &	-50:53:14 &	--  &	-68 &	-49 &	-37 &	--   &	15.5 &	-- \\
    \hline
  \end{tabular}
\end{table*}

\begin{table}
  \caption{Line fluxes for outflows and core, and their $1 \sigma$ errors, $T_{\rm MB} \, \Delta v$, in K\,\kms\ (i.e. corrected for beam efficiency), for the three CO isotopologues measured in each source (1).  For the \ce{^{12}CO} line the fluxes for the blue (2) and red-shifted (4) outflows are determined over the velocity ranges defined in Table~\ref{tab:outflowproperties}.  The line core velocity range used for the fluxes of \ce{^{12}CO} (3), \ce{^{13}CO} (5) and \ce{C^{18}O} (6) lines is between $v_{\rm B_2}$ and $v_{\rm R_1}$ in Table~\ref{tab:outflowproperties}. For each flux the corresponding error is listed. The errors include the random errors and an estimate in the uncertainty for the continuum level.}
  \label{tab:linefluxes}   
  \begin{tabular}{@{}cccccc}
    \hline
     & \ce{^{12}CO} & \ce{^{12}CO} & \ce{^{12}CO} & \ce{^{13}CO} & \ce{C^{18}O} \\
    Source & Blue & Core & Red & Core & Core \\
    (1) & (2) & (3) & (4) & (5) & (6) \\
    \hline
    G333.6--0.2	& 41 $\pm$ 3 & 218 $\pm$ 4 & 42 $\pm$ 2 & 66 $\pm$ 3 & 15 $\pm$ 2 \\
    G333.1--0.4	& 42 $\pm$ 3 & 369 $\pm$ 5 & 15 $\pm$ 2 & 102 $\pm$ 3 & 22 $\pm$ 2 \\
    G332.8--0.5	& -- $\pm$ 2 & 174 $\pm$ 4 & 39 $\pm$ 3 & 68 $\pm$ 3 & 14 $\pm$ 2 \\
    \hline 
  \end{tabular} 
\end{table} 

\begin{table}
  \caption{Fluxes for non-CO lines in Table~\ref{tab:obsdetails} together with their $1 \sigma$ errors, $T_{\rm MB}\, \Delta v$, in K\,\kms\ (i.e. corrected for beam efficiency), measured in the core component of each source. The errors include the random errors and an estimate in the uncertainty for the continuum level.  The velocity range used is between $v_{\rm B_2}$ and $v_{\rm R_1}$ in Table~\ref{tab:outflowproperties}. }
  \label{tab:otherfluxes}   
  \begin{tabular}{@{}cccccc}
    \hline
    Source & CS & \ce{HCO+} & \ce{H^{13}CO+} & SiO & \ce{N_2H+} \\
    \hline
    G333.6--0.2	& 32$\pm$2 & 19$\pm$2 & 6.3$\pm$3.1 & 1.7$\pm$1.3 & 8.3$\pm$1.3 \\
    G333.1--0.4	& 53$\pm$2 & 35$\pm$1 & 8.4$\pm$2.5 & 4.2$\pm$1.3 & 19$\pm$3.7 \\
    G332.8--0.5	& 24$\pm$2 & 15$\pm$1 & 4.2$\pm$1.5 & 1.3$\pm$1.1 & 14$\pm$1.4 \\
    \hline 
  \end{tabular} 
\end{table}

We defined the systemic velocity of the outflow sources and their core/outflow velocity boundaries by fitting 1-dimensional Gaussians to the \ce{C^{18}O}, \ce{^{13}CO} and \ce{CS} line profiles taken at the positions along the line-of-sight to the outflow centres. These lines appear to trace the core features only and have little emission in the wings. The systemic and FWHM velocities so determined are listed in Table~\ref{tab:gaussfits}. It can be seen that for each source the $v_{\rm sys}$ measurements agree to within 1~\kms. In addition, the \ce{^{13}CO} and CS lines are seen to be moderately optically thick, evident by the $\sim 1$~\kms\ larger values for their FWHM. 

The CS emission in the core of G333.1--0.4 (Figure~\ref{fig:profiles}) appears to be comprised of two blended velocity components which could not be fit with a single Gaussian.  The \ce{C^{18}O} emission has a smoother core profile and provided a better fit in this source. For consistency in our analysis, we therefore use the parameters determined from the fits to the \ce{C^{18}O} profiles in the calculations below. We measured the outflow masses and energetics using the observed brightness ratios between the \ce{^{12}CO}, \ce{^{13}CO} and \ce{C^{18}O} line emission.  Where these lines are well detected ($\rm S/N > 5$) we may use the ratio with the main line to determine the optical depth of the emission.  We summarise here our use of a radiative transfer analysis of the observed emission.

The radiative transfer equation has the general solution for the intensity of a velocity channel in a spectral line, known as the Detection Equation \citep[e.g.][Appendix C]{Stahler2005}:

\begin{equation}\label{eqn:det}
  T = T_{\rm A}^* / \eta_{\nu} = f [J_{\nu}(T_{\rm ex}) - J_{\nu}(T_{\rm BG})] (1 - e^{-\tau_{\nu}})
\end{equation}

\noindent where $T_{\rm A}^*$ is the measured intensity and $\eta_{\nu}$ the beam efficiency (taken as 0.55 from \citet{Ladd2005}).  $f$ is the beam filling factor for the emission, $J_{\nu}(T) = [h \nu / k ]/[e^{(h \nu / k T)} -1]$ with $T_{\rm ex}$ being the excitation temperature and $T_{\rm BG}$ being the temperature of the cosmic background radiation (i.e. 2.726 K).  $\tau_{\nu}$ is the optical depth of the emission.

Generally, in using Equation \ref{eqn:det}, we assume for the line isotopologue pair, say \ce{^{12}CO} and \ce{^{13}CO}, that the former is optically thick, while the latter is optically thin every where. However \ce{^{13}CO} unlikely to be optically thin in the cores of these sources, and thus the derived core masses are a lower limit. We also assume they have the same excitation temperature and beam filling factor. Furthermore, $\tau_{\ce{^{12}CO}}$/$\tau_{\ce{^{13}CO}} = N(\ce{^{12}CO})/N(\ce{^{13}CO}) = X[\ce{^{12}C}/\ce{^{13}C}]$, the abundance ratio of \ce{^{12}CO} to \ce{^{13}CO}. We have adopted abundance ratios of $X[\ce{^{12}C}/\ce{^{13}C}] = 20$, as preliminary generic radiative transfer models undertaken as per \citet{Carolan2008} did not support a higher ratio, if the observed line intensities were to be reproduced. The follow-up modelling tailored to each of the sources is discussed in the companion paper \citep{Wiles2015} to this one mentioned in the Introduction (Section \ref{sec:intro}). For the oxygen isotopes, we adopt a ratio of $X[\ce{^{16}O}/\ce{^{18}O}] = 100$. These values are smaller than the broad Galactic scale values \citep{Milam2005}, but taken into account of fractionation observed in cold dense core \citep{Mladenovic2014} and are consistent with \citet{Carolan2008} and \citet{Wiles2015}. Assuming these values, we see from the Detection Equation that $\tau_{\ce{^{13}CO}} = T_{\rm A}^*(\ce{^{13}CO})/T_{\rm A}^*(\ce{^{12}CO})$, with $\tau_{\ce{^{12}CO}}$ then determined by the preceding formula (and similarly for $\tau_{\ce C^{18}O}$).

We hence apply this analysis to determine the optical depth of the three \ce{CO} isotopologues lines, as a function of velocity. Furthermore, from standard molecular radiative transfer theory \citep*[e.g.][]{Goldsmith1999} we may show that the column density of the upper level of each transition is given by

\begin{equation}
  N_u = \frac{8 \pi k \nu^2}{A h c^3} T_{\rm A} \frac{\tau}{1 - e^{-\tau}} \Delta v
\end{equation}

\noindent where $k$, $h$ and $c$ are the well-known physical constants, $\nu$ is the frequency of the transition, $A$ is the radiative decay rate, and $\Delta v$ is the channel velocity spacing. The optically thin case ($\tau << 1$) simply has the optical depth correction factor $\tau/(1 - e^{-\tau})$ set to unity.

Once $N_u$ has been determined the total molecular column density can be found by applying $N_{\rm tot} = (N_u/g_u) Q(T_{\rm ex}) e^{E_u/kT_{\rm ex}}$ where $g_u = 2J+1 = 3$ is the level degeneracy and the partition function is $Q(T_{\rm ex}) = 2 T_{\rm ex}/T_0$ (with $T_0 = h \nu / k$) at an assumed excitation temperature, $T_{\rm ex}$.  We take the values of $T_{\rm ex}$ given in Table~\ref{tab:lineparams}, which are the peak values of $T_{\rm A}^*/\eta_\nu$ measured for the \ce{^{12}CO} line in each source. The observed \ce{^{12}CO} lines are extremely optically thick, and it is possible that the actual $T_{\rm ex}$ is higher than the temperature derived from the \ce{^{12}CO} flux, thus underestimating the total molecular column density. For a $T_{\rm ex}$ of 60 K, the total column density would be almost double that of an excitation temperature of 30 K.

In Table~\ref{tab:lineparams} we also list the maximum value of the optical depth found in each source, which corresponds to the minimum value of the \ce{^{12}CO}/\ce{C^{18}O} ratio, and the corresponding velocity. We also list the maximum values of \ce{^{12}CO}/\ce{^{13}CO} and \ce{^{12}CO}/\ce{C^{18}O} found for each source and their corresponding velocity.  These are seen to be very much less than the assumed abundance ratio, supporting the assumptions made above regarding the determination of the optical depth\footnote{Note also that any parameters derived from the line ratios are only calculated when the S/N in both corresponding lines is $> 5$.}.

\subsection{Molecular Lines: outflows, infall and turbulence} \label{sec:mol_in_out_turb}

\begin{figure}
  \includegraphics{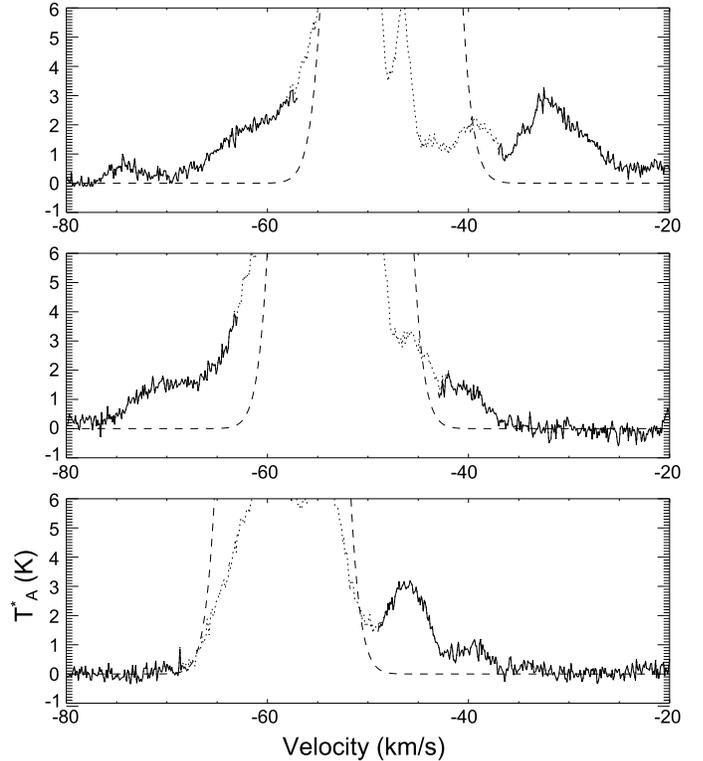}
  \caption{Optical depth-corrected, subtracted profiles, showing the outflow wings for each source. From top to bottom are shown G333.6--0.2, G333.1--0.4 and G332.8--0.5, respectively. The dotted line shows the \ce{^{12}CO} line profile and the dashed line the Gaussian fit to the \ce{C^{18}O} profile, respectively, scaled to the peak intensity of the optical-depth corrected \ce{^{12}CO} line.  The solid line shows the optical-depth corrected \ce{^{12}CO} profile with this scaled fit subtracted off (note that in the line core, where this subtraction is imperfect due to the scaling, it has been blanked out). This line represents the outflow profile. The y-axis scale is $T_{\rm A}^*$ in K and the x-axis is the $v_{\rm lsr}$ velocity in \kms.}
  \label{fig:tausubprofiles}
\end{figure}

In order to determine the velocity extent of the outflow, we use the optical depth of \ce{^{12}CO} (as calculated in Equation \ref{eqn:det}) to plot a core Gaussian line shape from \ce{C^{18}O}, and then from it to identify the outflow channels. The remaining profile (see Figure~\ref{fig:tausubprofiles}) shows the presence of extended red- and blue-wings in each source (except G332.8--0.5 for which no blue wing is evident).  We use these profiles to determine the velocity extent of the outflows, as listed in Table~\ref{tab:outflowproperties}, as well as the line core, for each source.  Table~\ref{tab:linefluxes} lists the CO line fluxes that have then been determined over these velocity ranges, corrected for the beam efficiency.  Table~\ref{tab:otherfluxes} provides the fluxes for the other lines observed in the line core velocity range.  In Table~\ref{tab:colden} the corresponding column densities are listed, calculated according to method described above using the CO isotopologues.  The Table includes the upper state column density, $N_u$, for the \ce{^{12}CO} (i.e.\ $J=1$) and the total \ce{H2} column density, $N_{\rm total}$.  $N_{\rm total}$ includes a correction for the optical depth listed in the Table (based on the \ce{C^{18}O} line) for the integrated flux in the line core, but assumes optically thin emission for the outflow wings (the \ce{^{13}CO} and \ce{C^{18}O} lines are not detected with sufficient S/N in the outflow wings to determine their optical depth).  A \ce{^{12}CO}/\ce{H2} abundance of $1.5 \times 10^{-5}$ is also assumed.

The molecular mass is given by $M_{\rm H_2} = N_{\rm total} \Omega d^2 \mu m_{\rm H_2}$ where $\Omega$ is the solid angle (for which the Mopra beamsize of $36\arcsec$ is used), the source distance $d = 3600$\,pc, $m_{\rm H_2}$ is the mass of a hydrogen molecule and $\mu = 1.2$ is the factor taken as the mean mass per hydrogen molecule (for fully molecular gas and helium mass fraction 10 per cent). We estimate that the ubiquitous low-level ambient emission may add an error of up to 10 per cent to the measured flux densities and hence the masses. A larger possible source of error lies in the probable blending of features that is impossible to disentangle without higher excitation transitions and/or higher angular resolution. 

We measure a mean length scale $l$ for the outflows by calculating the offset of the peak pixels in the extreme velocity channels from the centre position of the outflow. We then derive a mean outflow time scale $t_{\rm outflow}$ given by $t_{\rm outflow}= 2l / (v_{\rm B} + v_{\rm R})$. Finally, we calculated the quantities that characterise the direct mechanical feedback effects of the outflows, specifically mass loss rate $\dot{M}_{\rm loss}$, momentum $p$, mechanical force $F_{\rm mech}$, mechanical power $L_{\rm mech}$, kinetic energy $E_{\rm K}$ and free-fall time $t_{\rm ff}$:

\begin{equation}
  \dot{M}_{\rm loss} = \frac{M_{\rm B} + M_{\rm R}} {t_{\rm outflow}}
\end{equation}

\begin{equation}
  p = |M_{\rm B} v_{\rm B}| + |M_{\rm R} v_{\rm R}|
\end{equation}

\begin{equation}
  F_{\rm mech} = \frac{p} {t_{\rm outflow}}
\end{equation}

\begin{equation}
  E_{\rm K} = \frac{1}{2} (M_{\rm B} v_{\rm B}^2 + M_{\rm R} v_{\rm R}^2)
\end{equation}

\begin{equation}
  L_{\rm mech} = \frac{E_{\rm K}} {{t_{\rm outflow}}}
\end{equation}

\begin{equation} \label{eqn:t_ff}
  t_{\rm ff} = \sqrt{3 \pi / 32 G \rho}
\end{equation}

$t_{\rm ff}$ requires the average density $\rho$, determined from the average number density $n$ calculated as described in the next section (Section \ref{sec:fir}). The results of these calculations for the outflow parameters are given in Table~\ref{tab:derivedproperties}, together with the dust-derived mass and luminosities (also taken from Section \ref{sec:fir}).

\begin{table*}
  \caption{Derived column densities and masses. Upper level column density (\ce{^{12}CO} $J=1$) (3), (4) \& (5), total \ce{H2} column density (6), (7) \& (8) (assuming a [\ce{^{12}CO}/\ce{H2}] abundance of $1.5 \times 10^{-5}$) and \ce{H2} mass in the blue and red-shifted outflows and in the core component (9), (10) \& (11), for the three sources (1). For the core component the mean optical depth (2), determined from the \ce{^{12}CO}/\ce{C^{18}O} flux ratio, is used to correct for extinction in the determination of $N_{\rm total}$ and the mass.  The values listed for the outflow components (and $N_{\rm upper}$ for the core), do not have any extinction correction applied; i.e.\ optically thin emission is assumed for the outflow parameters. The upper limits listed for the blue outflow for G332.8--0.5 assume the $1 \sigma$ error for the flux given in Table~\ref{tab:linefluxes}.}
  \label{tab:colden}
  \begin{tabular}{@{}ccccccccccc}
    \hline
    Source & Mean $\tau$ & \multicolumn{3}{c}{$N_{\rm upper}$ ($10^{16} \times$ cm$^{-2})$} & \multicolumn{3}{c}{$N_{\rm total}$ ($10^{21} \times$ cm$^{-2})$} & \multicolumn{3}{c}{Mass (\solarM)} \\
     & Line Core & Blue & Core & Red & Blue & Core & Red & Blue & Core & Red \\
    (1) & (2) & (3) & (4) & (5) & (6) & (7) & (8) & (9) & (10) & (11) \\
    \hline
    G333.6--0.2 & 6.6 &	2.7 & 15 & 2.8 & 7.8 & 280 & 8.0 & 34 & 1200 & 34 \\
    G333.1--0.4 & 5.9 &	2.8 & 25 & 1.0 & 8.9 & 460 & 3.1 & 38 & 1970 & 14 \\
    G332.8--0.5 & 8.3 &	$<0.12$ & 12 & 2.7 & $<0.23$ & 190 & 5.0 & $<1$ & 810 & 22 \\
    \hline 
  \end{tabular} 
\end{table*}

\begin{table*}
  \centering
  \caption{Derived outflow parameters. The columns are as follows: (1) source; (2) \& (3) $M_{\rm B}$ \& $M_{\rm R}$ are the blue and red outflow masses respectively; (4) $M_{\rm core}$ molecular mass derived from the line core emission; (5) $l$ is the outflow length scale; (6) $t_{\rm outflow}$ outflow time for this scale size, given the outflow speeds listed in Table~\ref{tab:outflowproperties}; (7) $t_{\rm ff}$ the free-fall time; (8) $\dot{M}_{\rm loss}$ the mass loss rate derived for each outflow (red + blue components); (9) $p$ total outflow momentum; (10) $F_{\rm mech}$ the mechanical force for the outflows; (11) $L_{\rm mech}$ their mechanical luminosity and (12) $E_{\rm K}$ their kinetic energy.}
  \label{tab:derivedproperties}
  \scriptsize
  \begin{tabular}{@{}cccccccccccc}
    \hline
    Source & $M_{\rm B}$  & $M_{\rm R}$ & $M_{\rm core}$ & $l$ & $t_{\rm outflow}$ & $t_{\rm ff}$ & $\dot{M}_{\rm loss}$ & $p$ & $F_{\rm mech}$ & $L_{\rm mech}$ & $E_{\rm K}$ \\
     & \solarM & \solarM & \solarM & pc & $10^{3}$\,yrs & $10^{3}$\,yrs & $10^{-3}$ \solarM/yr & \solarM\kms\ & \solarM\kms\,yr$^{-1}$ & \solarL & $10^{47}$\,erg \\
    (1) & (2) & (3) & (4) & (5) & (6) & (7) & (8) & (9) & (10) & (11) & (12) \\
    \hline
    G333.6--0.2 & 34 & 34 & 1200 & 0.17 & 10 & 6.2 & 6.4 & 1100 & 0.10 & 130 & 1.7 \\
    G333.1--0.4 & 38 & 14 & 2000 & 0.50 & 40 & 13 & 1.5 & 800 & 0.023 & 29 & 1.2 \\
    G332.8--0.5 & $<1$ & 22 & 810 & 0.67$^*$ & 80$^*$ & 18 & 0.27$^*$ & 340 & 0.0040$^*$ & 5.2$^*$ & 0.53 \\
    \hline
  \end{tabular}
  \newline
  * since there is no blue outflow detected, the outflow length scale for this source is from the peak red outflow to the core centre, and this number is used for subsequent calculation.
\end{table*}

\normalsize

From Gaussian fits to the \ce{C^{18}O} and \ce{CS} line profiles, the cores have a velocity ($v_{\rm FWHM}$) of 6 - 7 \kms\ (Table \ref{tab:gaussfits}). Assuming a gas temperature of 20 K, the Mach number is above 12, while for 100 K gas, which is the warm component of dust temperature from SED fits (Table \ref{tab:sedfits}), the Mach number is 6. Thus, the turbulence of the cores are highly supersonic.

\subsection{Dust Continuum} \label{sec:fir}
Spectral energy distributions were fitted to the fluxes for the IR emission associated with the three sources (G333.6--0.2, G333.1--0.4 and G332.8--0.5). These fluxes were determined from a combination of {\it MSX} (8 - 21 \um), {\it IRAS} (25 - 100 \um), the Tata Institute of Fundamental Research (TIFR) balloon-borne telescope measurements at 150 and 210 \um\ \citep{Karnik2001} and 1.2-mm emission measured using SEST/SIMBA \citep{Mookerjea2004}, as listed in Table~\ref{tab:sed}. While the {\it IRAS} and balloon measurements used large apertures (3 arcminute for the balloon), inspection of the images across these wavebands showed they were dominated by a single source, at least within the 20 arcseconds resolution of the {\it MSX} and SIMBA data\footnote{Note however that for G333.6--0.2 the emission is somewhat more extended, so we also repeated the fitting for an aperture which includes all the emission within a 3 arcminute aperture.}. The application of the far-IR balloon data is particularly important in determining the source luminosity.  The fitting applied a 2-component greybody of the form

\begin{equation}
  F_{\nu} = \Omega_{\rm hot} B_{\nu}(T_{\rm hot}) + \Omega_{\rm warm} B_{\nu}(T_{\rm warm}) \epsilon_{\nu}
  \label{eqn:bbfit}
\end{equation}
 
\noindent
where the dust emissivity is given by $\epsilon_{\nu} = 1 - e^{-\tau_{\nu}}$, with the optical depth $\tau_{\nu} = \tau_0 (\nu/\nu_0)^{\beta}$, for a dust emissivity index $\beta$ taken to be equal to 2. $\tau_{\nu_o}$ corresponds to the wavelength $\lambda_0$ where the IR emission becomes optically thin (i.e.\ $\tau_{\nu_o} \equiv 1$). Fitting provides estimate for $T_{\rm hot}$ and $T_{\rm warm}$, representative temperatures for the hot and warm components of the fit, although only the warm component can be interpreted as physical parameter characteristic of the source \citep[see e.g.][]{Hill2009}. $B_{\nu}$ is the Planck black-body function. The angular sizes, $\Omega_{\rm hot}$ and $\Omega_{\rm warm}$, provide an effective source size for the IR emission, $r_{\rm hot}$ and $r_{\rm warm}$, at the distance to the source of 3.6~kpc. For reference, 0.1~pc corresponds to an angular size of 6 arcseconds, unresolvable with this data. The best fit parameters are listed in Table~\ref{tab:sedfits}.

The fitting also provides a source luminosity (the area under the spectral energy distribution), and a dust mass.  The latter is derived from the optically thin 1.2-mm (250-GHz) emission in conjunction with $T_{\rm warm}$, the dust temperature determined for the extended, warm component (which dominates the total flux) as follows: 

\begin{equation}
  M {\rm (dust)} = \frac{F_{\nu} D^2} {\kappa_{\nu} B_{\nu}(T_{\rm warm})}
\end{equation}

The total mass opacity coefficient was taken as $\kappa_{\rm 250\,GHz} = 0.005\,{\rm g}^{-1}\,{\rm cm}^2$ and the gas:dust mass ratio assumed to be 100. The core masses derived from the CO lines and the dust mass derived from the continuum are both given in Table~\ref{tab:derivedproperties}. As can be seen, the masses derived using the two methods are in good agreement given the assumptions made, differing by less than a factor of 3.

Luminosities are found to be 5 - 10\,$\times 10^5$ \solarL, dust masses a few thousand solar masses, and dust temperatures from 70 - 100\,K for the three sources. Note that including the extended emission around G333.6--0.2 nearly doubles the determined luminosity and dust mass from the region, but the other parameters determined are little changed. The dust luminosity to mass ratio is found to be $\rm \sim 400\, L_{\sun}/M_{\sun}$ in all cases. Given the fitted source size it is also possible to calculate the average density, column density and optical depth for each source, and these figures are also included in Table~\ref{tab:sedfits}. $M{\rm (dust)}$, $L_{\rm FIR}$ and $t_{\rm ff}$ (derived from $n$) are also included in Table~\ref{tab:derivedproperties}. Note that the column density, $N$ in Table~\ref{tab:sedfits} is applicable to the size of the warm component, $r_{\rm warm}$. This needs to be scaled to the 36 arcseconds Mopra beam size to be directly comparable to $N_{\rm total}$ in Table~\ref{tab:colden}, which requires multiplying $N$ by $\sim 0.02$, yielding values within a factor of 4 for the column density.

\begin{table*}
  \caption{IR source fluxes in Jy measured with {\it MSX}, {\it IRAS}, TIFR balloon-borne telescope \citep{Karnik2001} and SEST/SIMBA \citep{Mookerjea2004}.  The second row for G333.6--0.2 is for fluxes taken within a $3'$ aperture. This source is too confused to determine reliable {\it IRAS} fluxes in its large beam, so these are not included in the fits. Columns (1) and (2) provide our and Karnik's source ID; (3) \& (4) the source position; (5)--(8) the {\it MSX} fluxes; (9) -- (11) the {\it IRAS} fluxes; (12) \& (13) the Balloon fluxes; (14) the SEST/SIMBA flux.}
  \label{tab:sed}
  \begin{tabular}{@{}cccccccccccccc}
    \hline
    \multicolumn{2}{c}{IR Source} & \multicolumn{2}{c}{Position} & \multicolumn{4}{c}{{\it MSX}-band} & \multicolumn{3}{c}{{\it IRAS}-band} & \multicolumn{2}{c}{Balloon} & SIMBA \\
    Our ID & Karnik ID & RA & Dec & 8.3 & 12.1 & 14.6 & 21.3 & 25 & 60 & 100 & 150 & 210 & 1200 \\
     & & \multicolumn{2}{c}{J2000} & \multicolumn{10}{c}{\um} \\
    (1) & (2) & (3) & (4) & (5) & (6) & (7) & (8) & (9) & (10) & (11) & (12) & (13) & (14) \\
    \hline
    G333.6--0.2 & S23 & 16:22:10 & -50:06:00 & 461 & 3310 & 2970 & 6810 & & & & 27800 & 11600 & 82 \\
    ($3'$) & & & & 919 & 4479 & 4223 & 12050 & & & & 27800 & 11600 & 128 \\
    G333.1--0.4 & S15 & 16:21:01 & -50:35:15 & 133 & 378 & 512 & 1550 & 3040 & 12900 & 18400 & 20800 & 10000 & 34 \\
    G332.8--0.5 & S11 & 16:20:11 & -50:53:20 & 39 & 115 & 197 & 831 & 2030 & 14100 & 16800 & 15400 & 7790 & 20 \\
    \hline
  \end{tabular}
\end{table*}

\begin{table*}
  \caption{Derived source parameters using the fits to equation~\ref{eqn:bbfit}.  The second row for G333.6--0.2 is for the fluxes measured through a $3'$ aperture. Columns are as follows: (1) source ID; (2) \& (3) dust temperature for the hot and warm components; (4) cold temperature to which the temperature falls in the envelope at 1 pc; $T_{\rm cold}$ was estimated from a 1D radiative transfer model in {\sc dusty} \citep{ivezic&elitzur97} (5) \& (6) radii for the hot and warm components; (7) wavelength where the optical depth is unity; (8) source luminosity; (9) dust mass (converted into total gas mass); (10) average density; (11) column density (see comment in text for comparison to Table~\ref{tab:colden}); (12) visual extinction; (13) luminosity / mass ratio. }
  \label{tab:sedfits}
  \begin{tabular}{@{}ccccccccccccc}
    \hline
    ID & $T_{\rm hot}$ & $T_{\rm warm}$ & $T_{\rm cold}$ & $r_{\rm hot}$ & $r_{\rm warm}$ & $\lambda_0$ & $L_{\rm FIR}$ & $M{\rm (dust)}$ & $n$ & $N$ & $A_v$ & $L/M$ \\
     & K & K & K & $10^{-3}$\,pc & pc & \um & $10^6$\,\solarL & \solarM & $10^6$\,cm$^{-3}$ & $10^{24}$\,cm$^{-2}$ & mags & \solarL\,\solarM$^{-1}$ \\
    (1) & (2) & (3) & (4) & (5) & (6) & (7) & (8) & (9) & (10) & (11) & (12) & (13) \\
    \hline
    G333.6--0.2 & 350 & 98 & 20 & 2.1 & 0.081 & 752 & 1.2 & 3300 & 29 & 20 & 9800 & 370 \\
    (3') & 350 & 98 & 20 & 3.3 & 0.10 & 752 & 2.0 & 5200 & 23 & 20 & 9800 & 390 \\
    G333.1--0.4 & 240 & 69 & 20 & 3.3 & 0.11 & 371 & 0.57 & 1800 & 6.1 & 5.6 & 2800 & 310 \\
    G332.8--0.5 & 238 & 68 & 20 & 1.8 & 0.11 & 285 & 0.50 & 1100 & 3.5 & 3.3 & 1700 & 460 \\
    \hline
  \end{tabular}
\end{table*}

\subsection{Luminosities and ionizing photons} \label{sec:ionizing}
\citet{Fujiyoshi2006} used hydrogen radio recombination lines and the Bracket $\gamma$ IR line to derive the electron temperature and so calculate the number of Lyman continuum photons in G333.6--0.2. They estimate this to be $\rm \sim 9.5 \times 10^{49} \, s^{-1}$, coming from the inner $\sim 50$ arcseconds, equivalent to $\sim\,19$ O7V stars from their estimation of the number of Lyman continuum photons, assuming $\sim 5 \times 10^{48}$ s$^{-1}$ for one O7V star \citep{Martins2005}. 

\citet{Conti2004} estimated an IR luminosity of $1.1 \times 10^6$ \solarL with an aperture of 4 arcminute on the {\it MSX} and {\it IRAS} images, compare to $1.1 \times 10^6$ \solarL from our SED fittings with {\it MSX}, TIFR balloon and SIMBA data, and $2.0 \times 10^6$ \solarL\, if we take a 3 arcminute aperture (Table \ref{tab:sedfits}). Both of \citet{Conti2004} and our estimations of luminosity are near double  the luminosity from O4V and O5V stars found by \citet{Grave2014}, consistent with the finding of \citet{Kumar2013} that these two stars account for at least half of the luminosity of the UCH{\sc ii} region.

\section{Summary}\label{sec:summary}

We have used extensive molecular line data obtained with the Mopra radio telescope to search for evidence of outflow and infall associated with massive star formation in the G333 giant molecular cloud complex. The complexity of such sources where widespread massive star formation is under way makes such searches difficult. Evidence is generally required from a variety of molecular species and their isotopologues to unravel the competing effects of multiple sources, optical depth and different evolutionary states within the region. We have used data on the 3mm-band emission from eight molecular species, including three isotopologues of CO, dense gas tracers such as CS and \ce{N2H+}, and shock tracers such as \ce{HCO+} and SiO, together with archival continuum data from four IR surveys ({\it Spitzer}/GLIMPSE, {\it MSX}, {\it IRAS}, TIFR balloon), for this purpose. The line data has 0.6 arcminute and 0.1 \kms\ spatial and spectral resolution, and we have used it to determine physical parameters for sources in G333 to characterise their properties.

We have identified three massive star forming sources within G333 showing evidence of both infall and outflow: G333.6--0.2, G333.1--0.4 and G332.8--0.5. Outflow is evident by the broad wings to some of the line profiles, and infall by line splitting. These three sources are at different evolutionary states, with G332.8--0.5 at the earliest stage of star formation with the IR sources and H{\sc ii} regions least prominent, and vice-versa for G333.6--0.2. G333.1--0.4 lies in-between. All sources show broad profiles characteristics of outflows. This is particularly prominent in the blue wings for G333.6--0.2 and G331.1--0.4, but only evident in the red-wing for G332.8--0.5; the blue-wing is absent in this latter source. A clear outflow axis can be defined between the offset red and blue lobes for the first two sources, though the outflow itself remains morphologically poorly defined, in contrast to outflows seen in typical low mass star forming regions. HCO$^+$ and SiO emission, prominent in shocks, is also detected in all three sources, consistent with outflows being present in them.

Infall signatures are also apparent in the form of a red-blue asymmetry in all three sources, the line being self-absorbed by a cold, central continuum source. G333.6--0.2 is the most extreme with a splitting of $\sim 15$ \kms, and G332.8--0.5 the least at $\sim 5$ \kms.

We have used the ratio of the three CO isotopologues to correct for optical depth at each velocity channel, and so determine the column density for the line core as well as the outflow lobes. From this we are able to determine their masses, as well as estimate mass loss rates, outflow mechanical energies and luminosities. Typical outflow masses are 10 to 40 \solarM in each lobe, compared to core masses of order $10^3$ \solarM. Outflow size scales are a few tenths of a parsec, timescales are several $\times 10^4$ years and mass loss rates a few $\times 10^{-4}$ \solarM/yr. Flow momenta are $\sim 1000$ \solarM\,\kms and their mechanical luminosities a few $\times 10$ \solarL.

The source SEDs were used to calculate their luminosities, then by fitting to a 2-component grey-body model also the dust mass, dust temperature and source size for the extended component. Luminosities are $\sim 10^6$ \solarL, dust masses a few $\times 10^3$ \solarM (similar to that inferred from the line emission), dust temperatures $\sim 100$\,K and sizes $\sim 0.1$\,pc. This yields number densities of a few $\rm \times 10^6\,cm^{-3}$ and luminosity/mass ratios, $\rm L/M \sim 400$ \solarL/\solarM. The dust luminosity is also similar to that inferred from the hydrogen $\rm Br \gamma$ flux for G333.6--0.2.

Parameters for the infall may also be inferred from the line splitting in the profiles. However this is a complex procedure, as the radiative transfer needs to be considered. Estimates cannot simply be made from the magnitude of the line splitting, but must consider the medium through which the radiation passes.  This requires a more sophisticated approach than that presented here, and involves modelling of the source geometry as well as its physical characteristics in order to yield line profiles. In a companion paper \citep{Wiles2015} we apply such a 3D radiative transfer analysis making use of the code {\sc MOLLIE}, which is able to consider the competing contributions of the outflow, infall and ambient gas, which may also have different densities, temperatures and chemical compositions, in order to provide an estimate of the source parameters, and in particular to yield mass infall rates and infall speeds from the data set.

\section*{ACKNOWLEDGEMENTS}
We would like to thank the anonymous reviewer on improving this paper. NL's postdoctoral fellowship is supported by a CONICYT/FONDECYT postdoctorado, under project no. 3130540. NL acknowledges partial support from the ALMA-CONICYT Fund for the Development of Chilean Astronomy Project 31090013, Center of Excellence in Astrophysics and Associated Technologies (PFB 06) and Centro de Astrof\'{i}sica FONDAP\,15010003. MPR acknowledges funding from a Science Foundation Ireland grant 06/RFP/PHY051. LB acknowledges support from CONICYT Project PFB 06. The Mopra Telescope and ATCA are part of the Australia Telescope and are funded by the Commonwealth of Australia for operation as National Facility managed by CSIRO. The UNSW-MOPS Digital Filter Bank used for the observations with the Mopra Telescope was provided with support from the University of New South Wales, Monash University, University of Sydney and Australian Research Council. This work has made use of the SIMBAD database, operated at CDS, Strasbourg, France and also NASA's Astrophysics Data System. We have also made use of the NIST Recommended Rest Frequencies for Observed Interstellar Molecular Microwave Transitions, by Frank J. Lovas. We thank the referee for the constructive comments on improving this work.

\bibliographystyle{mn2e}
\bibliography{outflows_obs}


\bsp

\label{lastpage}

\end{document}